\documentclass[twocolumn,showpacs,preprintnumbers,amsmath,amssymb]{revtex4}

\usepackage{graphicx}
\usepackage{dcolumn}
\usepackage{bm}

\begin{document}

\preprint{Theoretical and Mathematical Physics, 139 (3): 787-800
(2004)}

\title{THE DARBOUX TRANSFORMATION AND EXACTLY SOLVABLE
COSMOLOGICAL MODELS}

\author{A.V. Yurov}
\email{artyom_yurov@mail.ru}
\author{S. D. Vereshchagin}%
 \email{sergev@nightmail.ru}
 \affiliation{%
The Theoretical Physics Department, Kaliningrad State
University,A. Nevskogo str., 14, 236041,
 Russia.
\\
}%


\date{\today}

\begin{abstract}
We present a simple and effective method for constructing exactly
solvable cosmological models containing inflation with exit. This
method does not involve any parameter fitting. We discuss the
problems arising with solutions that violate the weak energy
condition.
\end{abstract}

\maketitle

\section{\label{sec:level1}Introduction
} Scalar fields occupy a central place in modern cosmology.On one
hand, these fields (or field) play the role of inflatons, i.e.,
fields responsible for the inflation in the early universe
\cite{1}-\cite{3}. On the other hand, massless scalar fields are
currently considered a quintessence, a physical substratum
responsible for the observed accelerated expansion of the universe
\cite{4}, \cite{5}.

According to the most advanced cosmological model today, the
so-called chaotic inflation theory \cite{3}, it is assumed that
the early universe was dominated by a scalar field with minimal
coupling (in the simplest case) and that the energy density was
concentrated in the self-action potential. Under these conditions,
the initial equations can be considerably simplified. After the
simplification, the obtained system is easily integrated, showing
the existence of an inflationary phase (more accurately, a
quasi-de Sitter phase during which the Hubble parameter can be
considered constant). It is then assumed that during the inflation
process, the kinetic term increases until the slow rolling
approximation is no longer applicable. It then follows that
inflation ends spontaneously. The next phase is the oscillation
phase, which is necessary to fill the universe, "emptied by
inflation," with elementary particles. According to the discussed
paradigm, all these particles were produced from the vacuum by a
rapidly oscillating scalar field whose oscillation amplitude
decreased with time. A secondary reheating occurred, after which
the universe not only was homogenous and flat but also was filled
with hot matter. In other words, at this point, we can use
Friedmann's cosmological equations for matter with a state
equation characteristic of an electromagnetic field. The
subsequent evolution of the universe follows the classical
scenario: the universe cools as it expands, radiation decouples
from matter at some point, and the universe then expands according
to Friedmann's "two-thirds" law, and so on.

The discovery of the current accelerated expansion of the universe
[4] was unexpected because this fact did not follow at all from
the existing inflation paradigm. The explanation demanded
additional hypotheses and assumptions about a quintessence
(associated with a "dark energy," which is again "overabundant" in
the universe) that is the cause of the universe's accelerated
expansion. The need for additional assumptions of course
contradicts the established scenario. Perhaps, the best that can
be done to preserve the current paradigm is to prove that both the
early inflation and the current acceleration are caused by the
same phenomenon, i.e., the same scalar field. If this proves
impossible, then the inflation scenario will become less
attractive because it will turn out that inflation alone cannot
describe the observable structure of the universe and its
dynamics.

The difficulties that the inflation cosmology faced in explaining
the current accelerated expansion of the universe yielded new
cosmological scenarios, such as the ekpyrotic [6] and the cyclic
[7]. These scenarios arose from a set of ideas about "brane
worlds" [8], which in turn were initiated by research in the still
hypothetical M-theory. Unfortunately, research shows that brane
cosmologies encounter numerous problems, the best solution for
which is the hypothesis of inflation on the visible brane [9].
Thus, these models turn out to be no more than exotic variants of
the inflation cosmology.

It is clear from the above that inflation is still the best (i.e.,
the most natural and the most economical) hypothesis that explains
most, if not all, peculiarities of the observable universe. It is
therefore sensible to preserve this hypothesis without introducing
additional assumptions about the existence of a quintessence,
which would greatly weaken the status of inflation cosmology as a
descriptive physical theory.~\footnote{The alternatives to a
quintessence are models with a nonzero cosmological term.
Unfortunately, in terms of modern field theory(or string theory),
it is impossible to obtain a model with such a small vacuum energy
density that would not conflict with observations} For this, it is
reasonable to introduce the hypothetical inflaton as the cause of
the observed accelerated expansion of the universe. We note that
this idea is unpopular among cosmologists. Nonetheless, from our
standpoint, the proof of the nonequivalence of the inflaton and
the "dark energy" greatly undermines confidence in the whole
inflation scenario.Indeed, there are no experimental
justifications for the existence of the scalar inflaton, except
for the very existence of a homogenous,isotropic,and practically
flat universe. Until recently, it was thought that inflation
solves all major problems of cosmology, and this lent credence to
the inflation scenario. If it turns out that certain global
properties of the universe cannot be explained in terms of
inflation (for example, the current accelerated expansion), then
we greatly weaken the only (and indirect) argument that supports
inflation (see [10] for an excellent survey of the conceptual
problems of cosmology).

\section {Exactly solvable cosmological models}

An excellent approach to the explanation of accelerated expansion
in terms of inflation was given in [11], [12]. The offered models
admit a nonpositive definite self-action potential. This has an
interesting implication: even if the density of matter in the
universe is exactly equal to the critical density, a collapse
phase may follow an expansion phase, whereas expansion lasts
infinitely long if the potential is positive definite. In the
presence of a negative minimum, the current phase of accelerated
expansion occurs under a set of natural assumptions [11], [12].

The developed extended gauge theories of supergravity with N =2,
4, 8 allow hoping that the described models can have physical
meaning. The de Sitter solutions correspond to the extreme points
of the effective potential, and the squares of the masses of the
scalar fields m for $N \geq 2$ are quantized in units of the
Hubble constant $H_0$. If $m^2/H_0^2= O(1)$, then it is possible
to describe the current accelerated expansion of the universe
[12]. These scalar fields have exceedingly small mass ($m \sim
10^{-33} GeV$) and can significantly change the value of the
cosmological constant during the dustlike ("dark") matter era.
Simultaneously, they do not alter the standard predictions of the
inflation theory, because these fields are far from the potential
minimum in the early universe and "turn on" only when the Hubble
parameter becomes of the order of $|m|$. Finally, calculations
show that quantum corrections to m are extremely small.

Thus, as we have seen, inflation cosmology is rather viable. We
note that a general study of the dynamics of the universe for a
given self-action potential is an exceptionally difficult
mathematical problem. Therefore, much research into model
cosmologies where the equations can be solved exactly was done
during the last seven years [13]. The corresponding theories are
called exactly solvable cosmological models (ESCMs) or simply
exact cosmologies [13]. All ESCMs are based on the extraordinarily
wide gauge arbitrariness of the
Friedmann-Lemaitre-Robertson-Walker (FLRW) metric. If we consider
the general case of universes with such a metric and a self-acting
scalar field, then by setting the evolution of the scale
parameter, for example, we can calculate the self-action potential
that would lead to such an evolution. In other cases, the dynamics
of the field or of the Hubble parameter were fixed. In the latter
case, it was convenient to introduce a new independent variable,
the number $e$ of expansions of the universe. We note that the
apparently boundless arbitrariness is narrowed by the energy
conditions: the weak, the strong, or the dominant  condition,
depending on the context. Therefore, not every possible evolution
of the scale parameter, for instance, can occur. We study this
question in greater detail.

An obvious drawback of ESCMs are "pathological" self-action
potentials that most likely cannot be justified by elementary
particle physics. Nonetheless, the study of ESCMs can be very
important for inflation cosmology. Indeed, the chaotic inflation
theory assumes that practically any potential satisfying certain
conditions (see the introduction) leads to inflation with exit,
oscillations, transition to the Friedmann phase of radiation
dominance, and so on. If this is indeed so, then the pathological
form of the potentials normally appearing in ESCMs is
insignificant. We can gather statistics over many ESCMs and make a
well-founded claim about the reliability of the claims of the
chaotic inflation theory.

The investigations conducted in the works cited above led to the
following conclusions:
\newline
1.The idea of slow rolling is true. Inflation does indeed occur
under an extremely broad range of self-acting potentials, and
there is hence no need to fix a certain form of the potential to
obtain an inflationary universe.~\footnote{This was done in Guth's
early models, where the inflation was tied to the Higgs form of
the potential[1].}
\newline
2.The exit from inflation turned out to be a difficult problem.
For many model potentials, the universe never stops inflating. The
exit is generally achieved by fine-tuning or, to put it simply, by
parameter fitting.
\newline
3.The origin of oscillations is unclear. Even though exact models
of inflation have been constructed, they definitely do not contain
an oscillation phase (unlike nonintegrable models such as $V = m^2
\phi^2/2$, where the presence of oscillations is confirmed by
qualitative estimates and numerical integration [3], [14]). This
does not mean that such a state cannot be described by any exact
cosmology. For example, a potential that describes damped
oscillations of a scalar field was constructed in [15]. But it is
doubtful that such a potential can be obtained from quantum field
theory or string theory.
\newline
4. Every ESCM with exit has the following drawback: after the exit
from inflation, the universe either enters a radiation-dominated
stage, where it stays forever, or enters a dustlike matter stage
that is succeeded by a radiation-dominated stage, whereas it
should be the other way around. This is a fairly common
phenomenon. For example, an ESCM with a complex scalar field was
studied in [16], and the same problem occured.The most successful
ESCM model to date, described in [15], introduced an additional
hypothesis of the existence of a certain phase transition that
would deliver the universe from a radiation-dominated era to an
era dominated by dustlike matter. But a physical theory can hardly
be based on such ad hoc hypotheses. The last problem can seem
somewhat artificial. Indeed, according to the classical paradigm,
the transition from the radiation era to the dustlike matter era
is normally described by the Friedmann equations, the main
contributions to the energy-momentum tensor being from the
electromagnetic field (during the radiation era) and later, when
the temperature has fallen enough, from nonrelativistic matter
(hydrogen). But the difficulties that arise in item 4 have to do
with the solutions of the equations describing the scalar field
and gravity, and the scalar field during the decoupling stage is
very small compared to the radiation and baryons. Therefore, its
dynamics should not play a significant role.

Modern observational results do not agree with this conclusion. It
is now known that the main contribution to the energy-momentum
tensor in the current era comes from "dark energy." The observed
microlensing shows that regular baryonic matter is the next in
distribution after dark matter. In other words, even at the
current stage, the dynamics of the universe is governed not by
baryons but by "dark energy."

What is this "dark energy"? A reasonable assumption is that it is
described by a scalar field of unknown nature. It is also
reasonable to assume that "dark energy" has always played a
dominant role in the evolution of the universe. In other words,
not just inflation but indeed the entire post inflation dynamics
is governed a scalar field (or scalar fields). Taking our limited
knowledge of the universe into account, we naturally attempt to
describe cosmological evolution using only one scalar field.
Below, we discuss this specific problem.~\footnote{From this
standpoint, such elegant models as Chaplygin's gas model[17] are
unsatisfactory, if not necessarily untrue. The same can be said of
hybrid models[18].} Specifically, we propose a simple method for
constructing self-acting potentials that lead to inflation (or to
several inflations). In this approach, inflation with exit
versions arise without parameter fitting of any kind, containing
an asymptotic Freidmann era characteristic of dustlike matter.

This method has already been used in the theory of nonlinear
integrable equations and is known as the Darboux transformation
(DT) [19]-[21].~\footnote{In the context of nonlinear integrable
equations, it is proper to call it the Darboux-Crum-Matveev
method.} To avoid misunderstandings, we stress that the goal of
this paper is not to construct an adequate cosmological model or
to propose a hypothesis about the physical nature of "dark energy"
(as is done, e.g., in [12]) but to describe a mathematical method,
which, as we hope, can be used to construct such a model in the
future, if this is indeed possible.

\section {Formulation of the problem}

We consider the simplest case of a flat universe with the FLRW
metric
$$
ds^2=dt^2-a(t)^2\left (dx^2+dy^2+dz^2\right),
$$
filled with a scalar field with the Lagrange function
$$
L=\frac{1}{2}g^{\mu\nu}\partial_{\mu}\phi\partial_{\nu}\phi-V(\phi).
$$
Hence forth, we use the system of units $8\pi G=c=1$, commonly
accepted in cosmology. We limit ourselves to the flat case (k = 0)
because, on one hand, this is the simplest case and, on the other
hand, the curvature in the early universe is negligible. It is
easy to see that if $\phi=\phi(t)$, then Einstein's equations with
a cosmological term
$$
R_{\mu\nu}-\frac{1}{2}g_{\mu\nu}R=T_{\mu\nu}+g_{\mu\nu}\Lambda,
$$
reduce to the Schr\"{o}dinger equation
\begin{equation}
{\ddot\psi}=3\left(V+\Lambda\right)\psi, \label{1}
\end{equation}
where $\psi=a^3$. We assume here that $V=V(\phi(t))=V(t)$ and that
the time dependence of the field is given by
\begin{equation}
{\dot\phi}^2=-\frac{2}{3}\frac{d^2}{dt^2}\log\psi. \label{2}
\end{equation}
In this language, it is very easy to formulate a naive method for
constructing ESCMs. For example, we assign $\psi(t)$; substituting
in (\ref{2}), we determine the time dependence of the field
$\phi(t)$; and using (\ref{1}) and (\ref{2}) together, we find $V$
as a function of  $\phi$ in a parametric form.

The given method looks very simple. But it should not be thought
that such a potential can be found for {\em any} evolution of the
scale parameter (or the Hubble parameter $H={\dot a}/a$). First,
as is clear from (\ref{2}), a naive choice of $a(t)$ (i.e.,
$\psi$) most likely leads to situations where $\phi$ becomes
imaginary after some time. Second, only in exceptional cases is it
possible to reconstruct $V$ as an explicit function of $\phi$,
which of course hampers attempts to find a physical meaning of
such a potential. But in this paper, we concentrate on the first
point and disregard the second. As we see below, choosing $\phi$
such that the right-hand side of (\ref{2}) changes sign in a given
interval, we can construct cosmological models with inflation and
an inevitable exit from it.

In what follows, we require the energy conditions. Following [22],
for example, we define them as
\newline
\newline
1. the weak energy condition, $\rho\ge 0$, $p+\rho\ge 0$;
\newline
\newline
2. the strong energy condition, $p+\rho\ge 0$, $3p+\rho\ge 0$;
\newline
\newline
3. the dominant energy condition, $\rho\ge 0$, $\rho\ge
p\ge-\rho$,
\newline
where $\rho$ and $p$ are the respective density and pressure. It
is obvious that the weak energy condition (the most important one)
holds if and only if the right-hand side of (\ref{2}) is positive.
This simple property is very importance later.

We apply the DT [20], [21] to (\ref{1}). We have two
justifications for this. First, the DT is an isospectral symmetry
of the Schr\"{o}dinger equation because the DT preserves the
asymptotic characteristics of the transformed functions up to a
constant multiple. In particular, square-integrable functions stay
in that space if the support functions that generate the DT do not
have zeroes or have alternating zeroes if the number of functions
is even [23]. This property can be used as follows. Let the scale
parameter have an asymptotic characteristic of a flat universe
filled with dustlike matter, i.e., $a\sim t^{2/3}$. Dressing the
appropriate solution of Eq.(\ref{1}) with a support function from
$L_2$ (any regular solution of (\ref{1}) with the same potential
suffices, even if it does not belong to $L_2$), we obtain a new
solution of Einstein's equations with the same asymptotic
condition. In other words, even though the behavior of the scale
parameter can be rather odd (for example, it can grow rapidly on a
finite interval, and the universe therefore experiences an
inflationary period), the universe enters a Friedmann expansion
era with the two-thirds law in any case.~footnote{Of course, the
support function must have no zeros between the inflationary
era,where ${\ddot a}>0$ ,and the Friedmann era proper. Otherwise,
the inflationary era and the matter-dominated era are separated by
a singularity, i.e., represent two casually unconnected universes.
} This property guarantees an exit from inflation.

Second, DTs can be used to construct inflation models. For this,
we rewrite Eq.(\ref{2}) in the form ${\dot\phi}^2=V^{(1)}-V$,
where $V^{(1)}$ is the potential constructed from the potential
$V$ using Crum's formulas and the support function $\psi$ (see
formulas (\ref{5}) below) [21]. We assume that the function $\psi$
has $N$ zeroes.Then $V^{(1)}$ has at least $N$ singularities. We
assume that these singularities are not present in the initial
potential $V$. If $V^{(1)}$ is not positive definite in the
neighborhood of one of these singularities (or $V$ has a
singularity of its own, around which it is positive definite),
then there exist $t_1$ and $t_2$ on different sides of the
singularity such that ${\dot\phi(t_1)}= {\dot\phi(t_2)}=0$, for
the ${\dot\phi^2 (t)}<0,\,\,t_1<t<t_2$. The proof of this
statement is obvious. It is now clear that if the given conditions
are met, then the universe certainly enters an inflationary phase
when $t>t_1$.

\begin{figure}
\includegraphics{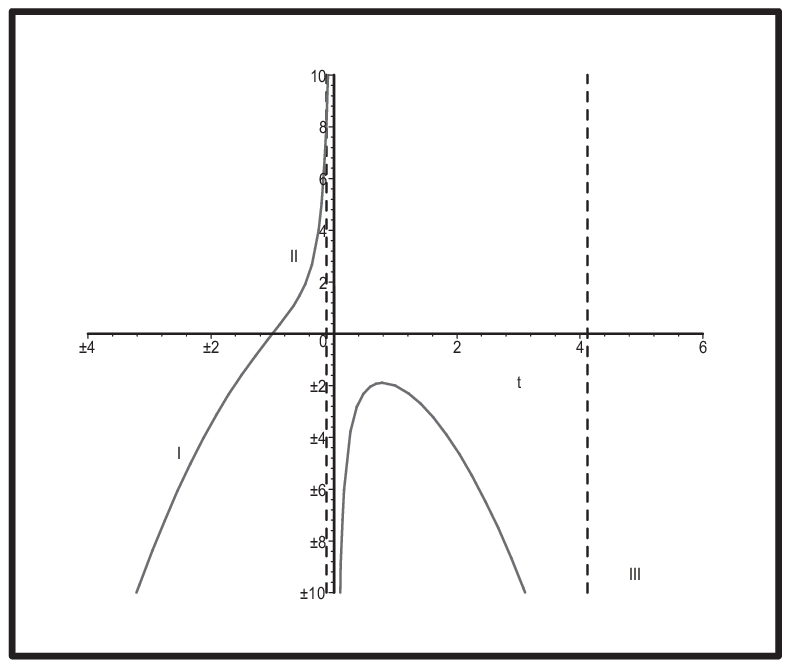}
\caption{\label{fig:epsart} }
\end{figure}

Indeed, we assume that the universe is born at $t=t_i$ ($t_i<t_1$)
and that there exists a period of time during which all energy
conditions are met. It can be shown that as we approach $t_1$, the
strong energy condition is violated first, while the weak one
holds. As a result, the universe expands up to the instant $t_1$.
At $t=t_1$, we obtain the de Sitter phase, but for only one
instant. A forbidden zone then follows,~\footnote{The existence of
a "forbidden" zone correlates with the known lack of
superinflation in cosmologies with scalar fields} which should be
removed by matching the solutions at $t=t_1$ and $t=t_2$. We note
that this matching requires stricter conditions than those found
in corresponding problems in nonrelativistic quantum mechanics
based on the Schr\"{o}dinger equation. Here, we require that not
only the wave functions and their first derivatives but also the
second derivatives be the same at the matching point. The latter
is necessary to avoid discontinuities of density and pressure. The
general solution of (\ref{1}) has two constants of integration;
therefore, it is not obvious that we can always satisfy all three
conditions. But it is easy to verify that the third condition
(equality of the second derivatives) follows automatically from
the first two! This fortunate circumstance is a consequence of
matching the solution at the inflection point of $\log\psi$. The
price we pay (a discontinuity of the third derivative of the scale
parameter) does not seem excessive, at least while we consider the
system as a classical (i.e., not quantum) system.

Thus, controlling the zeroes of the support function of the DT, we
can construct exact cosmologies with inflation. If solutions with
known asymptotic characteristics (for example, evolving according
to the two-thirds law) are used as initial base solutions, then
the universe necessarily exits inflation, without any parameter
fitting. Below, we demonstrate examples of how this approach
works.

\section {A simple model}

The law $a\sim t^{2/3}$ gives the potential
\begin{equation}
V=\frac{2}{3t^2}. \label{3}
\end{equation}
The general solution of (\ref{1}) with potential (\ref{3}) and
without a cosmological constant has the form
\begin{equation}
\psi^{(-)}=-\frac{c_1}{t}-c_2t^2, \label{4}
\end{equation}
where $c_1$ and $c_2$ are the constants of integration. The
solution describes two disconnected universes (Fig.1). The first
is infinite in the past but has a singularity at
$t_0=-(c_1/c_2)^{1/3}$. The second is rather strange. It is born
from a singularity at $t=t_0$, then expands, and undergoes
superinflation, i.e., the scale parameter becomes infinite at $t =
0$. Then the universe contracts, the scale parameter changing
sign. Of course, this solution cannot be taken literally. The
superinflation zone (to the left and right of $t=0$) is forbidden
because the weak energy condition is violated in it. It is easy to
find the density and pressure
$$
p=-\frac{c_1\left(c_1+4c_2t^3\right)}{t^2\left(c_1+c_2t^3\right)^2},\qquad
\rho=\frac{\left(2c_2t^3-c_1\right)^2}{3t^2\left(c_1+c_2t^3\right)^2}
$$
Therefore, the above mentioned forbidden zone is in the interval
between $t_1$ and $t_2$, where
\begin{equation}
\begin{array}{c}
\displaystyle{t_1=\left[\left(2-\frac{3}{\sqrt{2}}\right)\frac{c_1}{c_2}\right]^{1/3}
\sim -0.5 \left(\frac{c_1}{c_2}\right)^{1/3}},\\
\\
\displaystyle{t_2=\left[\left(2+\frac{3}{\sqrt{2}}\right)^{1/3}\right]\sim1.6\left(\frac{c_1}{c_2}\right)^{1/3}}.\label{5}
\end{array}
\end{equation}
We consider solution (\ref{4}) to hold only for $t\leq t_1$,where
$t_1$ is given by (\ref{5}). To continue the solution into the
region $t>t_1$, we rewrite (\ref{4}) with different constants of
integration, denoted by $b_1$ and $b_2$:
\begin{equation}
\psi^{(+)}=\frac{b_1}{t}+b_2t^2. \label{6}
\end{equation}
For (6), the weak energy condition is violated in the interval
$(t'_1,t'_2)$ where the $t'_{1,2}$ are determined by expressions
(\ref{5}) with the substitutions $c_1\rightarrow b_1$ and
$c_2\rightarrow b_2$. The constants $b_1$ and $b_2$ should be
chosen such that they satisfy~\footnote{It is easy to verify that
(\ref{7}) does not hold if $b_1=c_1$ and $b_2=c_2$; therefore, we
must introduce the second solution, Eq.(\ref{6}).}
\begin{equation}
\psi^{(-)}(t_1)=\psi^{(+)}(t_2'),\qquad
\dot{\psi}^{(-)}(t_1)=\dot{\psi}^{(+)}(t_2').\label{7}
\end{equation}
It is noteworthy that another relation,
\begin{equation}
 \ddot{\psi}^{(-)}(t_1)=\ddot{\psi}^{(+)}(t_2').\label{8}
\end{equation}
is a consequence of (\ref{6}). It thus follows that when matching
$\psi^{(-)}(t_1)$ and $\psi^{(+)}(t_2')$, the density, pressure,
and rate of density variation are continuous!

An obvious deficiency of this solution is a jump in time: $t_1\neq
t_2'$. Therefore, the solution is not defined on the interval
$(t_1,t_2')$. But it is easy to move $t$ in $\psi^{(+)}$ and
$\psi^{(-)}$ such that the two conditions that
\newline
1. the initial singularity corresponds to time $t=0$ and
\newline
2. $t_1=t_2'$ with no jump in time.
\newline
are met. We skip the calculations and present only the final
answer:
\begin{figure}
\includegraphics{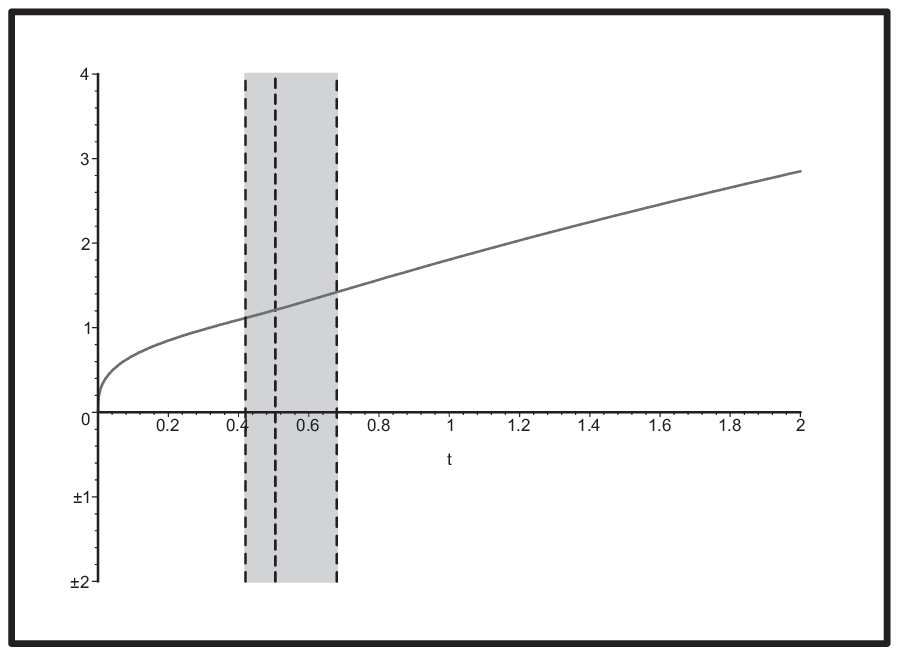}
\caption{\label{fig:epsart} }
\end{figure}


\begin{figure}
\includegraphics{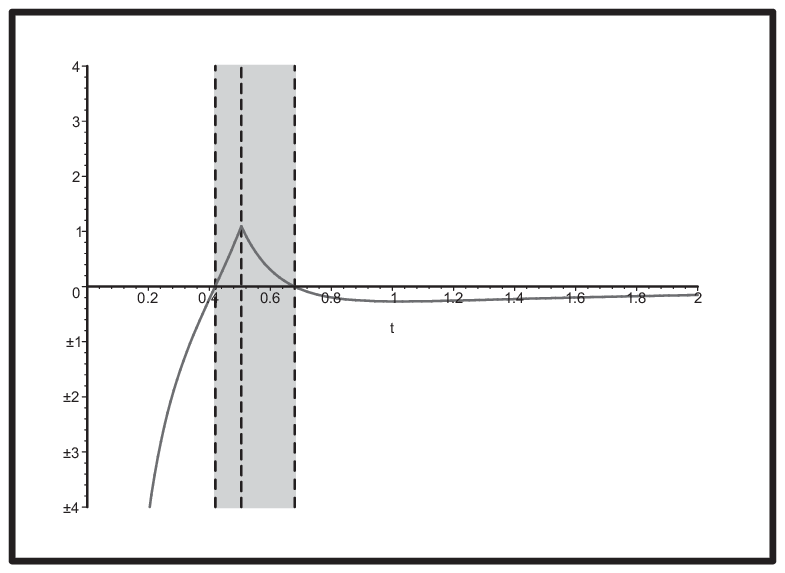}
\caption{\label{fig:epsart} }
\end{figure}


1. For $0<t\le t_1$, the dynamics are described by the formulas
\begin{equation}
\begin{array}{c}
\displaystyle{\psi^{(-)}(t)=-\frac{c_1^3}{t-c_1/c_2}-c_2^3\left(t-c_1/c_2\right)^2},\\
\\
\displaystyle{V^{(-)}=\frac{2/3}{\left(t-c_1/c_2\right)^2}}.\label{9}
\end{array}
\end{equation}
\newline
2.For $t_1<t$, the dynamics are described by the formulas
\begin{equation}
\begin{array}{c}
\displaystyle{\psi^{(+)}(t)=\frac{c_1^3(3-2\sqrt{2})}{t-g}+c_2^3(3+2\sqrt{2})\left(t-g\right)^2},\\
\\
\displaystyle{V^{(+)}=\frac{2/3}{\left(t-g\right)^2}},\label{10}
\end{array}
\end{equation}

Here,
\begin{equation}
\begin{array}{c}
\displaystyle{t_1=\frac{c_1}{c_2}\left(1+\left(2+\frac{3}{\sqrt{2}}\right)^{1/3}\right)\sim
2.6\frac{c_1}{c_2}}, \\
\\
\displaystyle{g=\frac{c_1}{c_2}\left(1+\left(16+12\sqrt{2}\right)^{1/3}\right)\sim
4.2\frac{c_1}{c_2}}.\label{11}
\end{array}
\end{equation}
we introduce the new constants $c^3_{1,2}$ for convenience, and
$t_2=t_1$ .The matching here is
$$\psi^{(-)}(t_1)=\psi^{(+)}(t_1),\qquad
{\dot\psi}^{(-)}(t_1)={\dot\psi}^{(+)}(t_1),$$
$${\ddot\psi}^{(-)}(t_1)={\ddot\psi}^{(+)}(t_1),\qquad
V^{(-)}(t_1)=V^{(+)}(t_1),$$
$$\rho^{(-)}(t_1)=\rho^{(+)}(t_1),\qquad
p^{(-)}(t_1)=p^{(+)}(t_1).$$

The graph of $a$ as a function of $t$ for the second universe is
given in Fig.2. The curve is described by the formula
$a^{(-)}=\psi^{(-)}(t)$ for $t<t_1$ and by $a^{(+)}=\psi^{(+)}(t)$
for $t>t_1$,where $\psi^{(-)}(t)$ and $\psi^{(+)}(t)$ are given by
(\ref{9}) and (\ref{10}). We specifically choose $c_1=c_2=1$.
Figure 3 describes the acceleration $\ddot{a}$ as a function of
time. The break is at $t=t_1$, given by (\ref{11}).

As can be seen, inflation occurs on a finite interval (where there
is a tooth on the $t$ axis).We also note that as
$t\rightarrow\infty$, the potential behaves as
$$
V\sim \frac{2}{3}e^{-\sqrt{3}\phi},\qquad \phi\sim
\frac{2}{\sqrt{3}}\log t.
$$

We have thus constructed a simple model of inflation with exit and
an asymptotic transition to the Friedmann expansion era, located
in the matter-dominated phase. Of course, this example does not
describe our universe, because there is no oscillation phase.
Neither is there secondary acceleration. But it is evident that
this method can be used without fundamental difficulties to
construct model cosmologies with exit without parameter fitting.
Neither it is necessary to use hybrid models. We note that the
solution describes another, collapsing universe with no beginning
but with an end. These solutions seem rather strange, but their
appearance is inevitable because the equations of general
relativity are invariant under time reversal.

\section{The Darboux transformation}

It is clear that any exact solution of (\ref{1}) satisfying the
energy conditions generates an exact solution of the Einstein
equations in the FLRW metric. Because not all solutions $\psi$ of
(\ref{1}) generate a positive definite expression after
substitution in the right-hand side of (\ref{2}), the exact
solutions of the Einstein equations are a subset of the exact
solutions of (\ref{1}). On the other hand, practically all exact
solutions of Schr\"{o}dinger equation (\ref{1}) with "physical"
potentials can be constructed using a DT~\footnote{We do not
consider potentials such as the rectangular well, limiting
ourselves to continuous functions.}. Examples are the harmonic
oscillator, multisoliton, and even finite-gap
potentials.~\footnote{A common idea for constructing many exactly
solvable cases is form-invariance. For example, the harmonic
oscillator and finite-gap potentials are constructed by closing a
chain of discrete symmetries. The same is true for certain
Painlev\'{e} transcendents.} We use this as an "experimental fact"
without attempting to give precise mathematical meaning to
statements like "all potentials for which Eq.(\ref{1}) can be
explicitly solved for all values of the spectral parameter are
constructed using the DT and, possibly, an additional finite set
of symmetries from other potentials with this property."
Informally speaking, we can then conclude that all exact solutions
of the Einstein equations in a flat FRLW universe filled with a
real scalar field can be constructed using DTs.

As a simple example, we consider potential (\ref{3}), which is
constructed using a single DT from a zero potential $V(0)= 0$. The
result of a DT applied $N$ times can be written compactly using
Crum's formulas:
$$
\psi^{(N)}=\frac{\Delta_{N+1}}{\Delta_N},\qquad
V^{(N)}=V-\frac{2}{3}\frac{d^2\ln\Delta_N}{dt^2},\label{12}
$$
where
$$
\Delta_N=det\left[\frac{d^k\psi_i}{d t^k}\right], \qquad
k=0,...,N-1;\qquad i=1,...,N,
$$
and
$$
\Delta_{N+1}=det\left[\frac{d^k\psi_i}{d t^k}\right],\qquad
k=0,...,N;\qquad i=1,...,N+1.
$$
In these expressions, the $\psi_i$ are particular solutions of
(\ref{1}) with a common initial potential $V$ and different
eigenvalues $3\Lambda$. It is now clear how we can use DTs to
construct models with two accelerations. We already mentioned that
the potential $V^{(N)}$must have singularities. It follows from
Crum's formulas that all singularities coincide with the zeroes of
functions that are contained in Crum's determinant $\Delta_N$
Therefore, the problem is reduced to controlling these zeroes. We
assume that the initial potential is given by (\ref{3}). Solution
(\ref{4}) corresponds to a trivial cosmological term. The
solutions of (\ref{1}) with positive ($\Lambda=k^2/3$) and
negative ($\Lambda=-k^2/3$) values of the lambda term are
\begin{equation}
\begin{array}{l}
\displaystyle{
\psi_+=\mu_1\left(1-\frac{1}{kt}\right)e^{kt}+\mu_2\left(1+\frac{1}{kt}\right)e^{-kt},}\\
\\
\displaystyle{
\psi_-=\nu_1\left(\frac{\sin\,kt}{kt}-\cos\,kt\right)+
\nu_2\left(\frac{\cos\,kt}{kt}+\sin\,kt\right),}
\end{array}\label{13}
\end{equation}
where $\mu_{1,2}$ and $\nu_{1,2}$ are arbitrary constants. Using
these functions for different values of $k$ in formulas (\ref{12})
and regarding $\psi_{_{N+1}}$ as defined by (\ref{4}) for all $N$,
we can construct a set of solutions with the Friedmann two-thirds
asymptotic behavior for $t\to+\infty$.

It is clear that the functions $\psi^{(N)}$ contain any number of
singularities near which the weak energy condition is violated and
acceleration therefore occurs. As in the example in Sec.3, the
solutions must be cut off at the inflection points $M_j$ of
$\log\psi^{(N)}$, where the exact de Sitter phase is realized, and
matched along these points. We do not need to match solutions at
adjacent points. By adjacent points, we mean points $M_j$ and
$M_{j+1}$ between which there is only one singularity of
$\psi^{(N)}$. As shown in Sec.3, we can always match any two
points $M_j$ , which gives a degree freedom in constructing the
required solution.

We note that the explicit form of the solution quickly becomes
very complicated as $N$ increases. We consider some examples.
Henceforth, we suppose that $c_1=c_2=1$ everywhere.

A single DT with a support function $\psi_+$ from (\ref{13}) gives
$$
\begin{array}{l}
\displaystyle{
\psi^{(1)}=\frac{\mu\left[\left(x-1\right)^3+k^3+1\right]e^{x}-
\left[\left(x+1\right)^3+k^3-1\right]e^{-x}}{\mu (x-1)e^{x}+(x+1) e^{-x}},}\\
\\
\displaystyle{V^{(1)}=\frac{2k^2}{3}\frac{\mu^2e^{2x}+e^{-2x}-4\mu
x^2-2\mu} {\left(\mu (x-1)e^{x}+(x+1) e^{-x}\right)^2},}
\end{array}
$$
where $\mu_1=1$, $\mu_2=\mu$, $x=kt$. For $\mu<0$ is given in
Fig.4. Here, we already removed the regions where the weak energy
condition is violated. The boundary points (where $\dot\phi=0$)
must be matched. It is then clear that such a solution describes
two universes. The first contracts to a final singularity,
experiencing a time-reversed inflation at a certain instant prior
to the final collapse. The dynamics of the second universe is a
time reversal of the evolution of the first: it is born from a
singularity, expands for some time, then undergoes inflation,
exits, and settles asymptotically on the Friedmann expansion,
characteristic of a matter-dominated universe.
\begin{figure}
\includegraphics{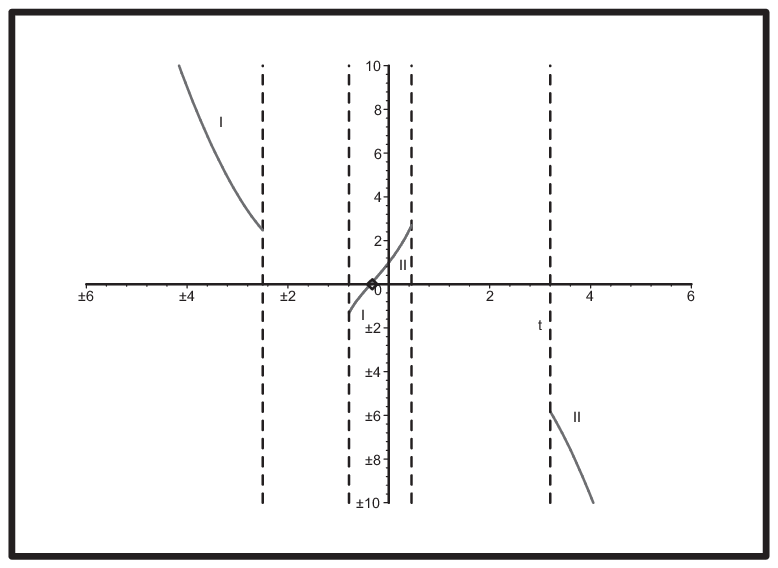}
\caption{\label{fig:epsart} }
\end{figure}


A new situation arises for $\mu>0$. Figure 5 shows the graph of
$\psi^{(1)}(t)$ for $\mu=2$. It demonstrates the evolution of
three universes. The first collapses into a finite singularity but
has no beginning. The second is born from an initial singularity,
expands for some time, then begins to contract, and collapses into
a final singularity (we recall that $k=0$). The third is born,
expands, undergoes inflation (the boundary points are matched),
and exits into a two-thirds phase.

As a final example, we present a solution obtained by a double DT
(Fig.6). One support function is a solution of (1) with
$\Lambda=+1/3$, $\mu_1=1$, $\mu_2=2$ and $k=1$; the other has
$\Lambda=+4/3$,  $\mu_1=1$, $\mu_2=3$ and $k=2$. We also have
three universes here. We note that the second exists for a finite
period of time, during which it undergoes inflation and then
collapses. These solutions are very similar to the solutions in
[11] that we mentioned at the beginning of Sec.2.

We have not attempted to construct models with two (or more)
accelerations. It is clear that this problem can be solved in
principle using this method. Moreover, the acceleration rate is
determined by the density $\rho$; therefore, constructing
potentials with several singularities, we can obtain several
inflations, each with its own intensity. If the second inflation
corresponds to a smaller value of $\rho$ than the first, then it
is much slower. But we think that it is too early to begin
constructing models using DTs. It is necessary to understand how
an intermediate oscillation phase can be included. It is still
unclear to us how can this be done. It is obvious that any
inflation model without such a phase is impracticable.~\footnote{A
certain hint is given by the astonishing potentials obtained by
closing the Veselov-Shabat chain[24]. These potentials contain
amplitude-modulated oscillations and must be considered exactly
solvable. The spectra comprise a set of several arithmetic
progressions, and depending on a special condition, the solutions
either can be expressed in terms of Riemann's $\theta$-function or
are generalizations of the Painlev\'{e} transcendents. In
particular, the solution of a three-link ($N=3$) chain can be
expressed in terms of the transcendents of the $P_{IV}$ equation.}
\begin{figure}
\includegraphics{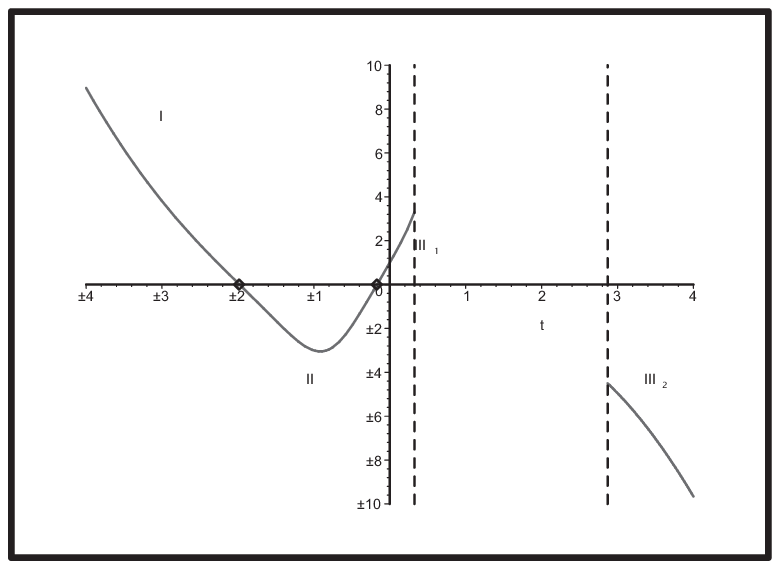}
\caption{\label{fig:epsart} }
\end{figure}


\begin{figure}
\includegraphics{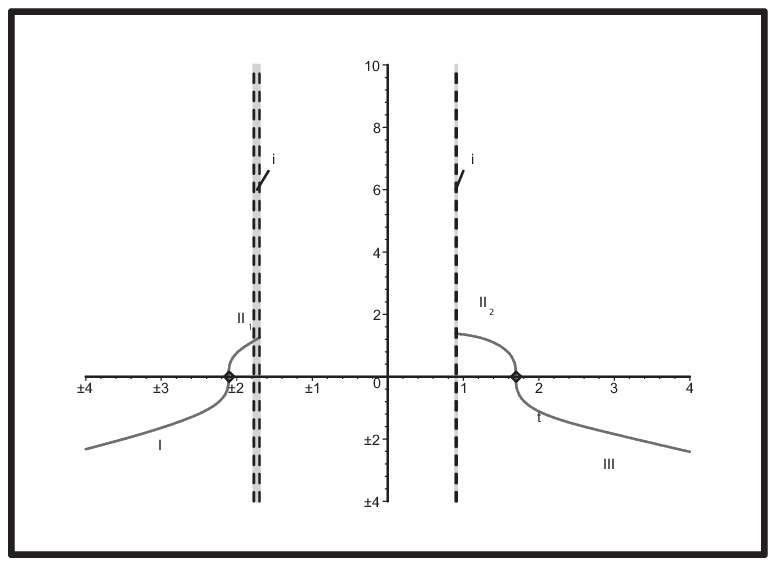}
\caption{\label{fig:epsart} }
\end{figure}


\section{Conclusion}

We suppose that the exact solution $a(t)={}^3\sqrt{\psi^{(N)}(t)}$
describing two acceleration phases and an intermediate oscillation
phase is constructed. We obviously encounter the problem of
interpreting the potential $V^{(N)}$, which is undoubtedly
hideous. It is unlikely that it can ever be described by any field
theory or even string theory. Indeed, even the potentials obtained
by a single DT are already too complicated and unphysical, and
they are simple enough to describe even two consecutive
accelerations without oscillations!

A natural reaction to the appearance of such expressions is the
conclusion that real-life cosmology is not exactly solvable. From
this standpoint, exact cosmologies can be useful models, but they
are incapable of describing the real universe. But just how
correct is this conclusion? It is quite possible that the correct
potential of a self-acting scalar field is not defined by
elementary particle physics but appears for a different reason.
Any suppositions of this kind are speculative and unreliable.
Nonetheless, we attempt to point out a source of such strange
potentials outside both field theory and string theory (of the
future).

According to the chaotic inflation theory of Linde, all possible
vacuum states are realized in different areas of the universe,
which in turn must lead to different mass spectra of the
elementary particles populating these regions. In this paradigm,
our region of the universe cannot be identified using physical
laws alone; a certain "weak anthropic principle" is also required.
In other words, even if an all-encompassing M -theory can be
constructed and it might justify our expectations, it would still
prove insufficient to explain why the vacuum state of our universe
is the way it is and not different, and so on.

To illustrate this picture, Linde introduces the concept of the
multiuniverse[14]. In the simplest case, the multiuniverse can be
described as an infinite sum over all possible actions describing
all possible quantum field theories and M -theories.~\footnote{In
fact, Linde goes even further, proposing to include even those
models with no description in the Lagrangian formalism and even
discrete models such as cellular automata. Some do not stop even
here, considering all possible mathematical structures to be
elements of the multiuniverse(see[26]).} Even though hypotheses of
this kind can seem speculative, it must be borne in mind that all
existing explanations of the observed density of the vacuum energy
($\sim 0.7\rho_c$) have been obtained using the anthropic
principle, which apparently makes sense only if the multiuniverse
hypothesis is true.~\footnote{An unusual alternative is the
"self-consistency" hypothesis proposed in [27].}

If we accept this hypothesis, then it is clear that the existence
of regions with the strangest self-acting potentials is possible.
In this case, we must readdress the basic questions we normally
ask. For example, potentials $V(\phi)$ are generally taken from
particle theory ($V\sim \phi^2$ and $V\sim \phi^4$) or string
theory $V\sim e^{\alpha\phi}$. Why these and not $V\sim\phi^8$,
for example? Because a theory with such a potential is
nonrenormalizable for $d=4$. We have every reason to believe that
a nonrenormalizable theory is a "bad" theory because it requires a
new Lagrangian on each new scale and therefore contains an
infinite number of indefinite parameters as a rule. As well as
being renormalizable, models must also satisfy other generally
known properties: unitary, a Hamiltonian bounded from below, etc.

We now suppose that we have succeeded in constructing a family of
potentials giving an evolution of the universe consistent with
observations (for example, using DTs). The next step would be to
check these models using the methods of quantum theory
(renormalizability, etc.). If one of these potentials satisfies
the appropriate checks, then we can perfectly well use it even
though we did not obtain it using the basic principles of field
theory or string theory. We obtained it using the multiuniverse
concept, and field theory (or string theory) is only required to
verify its consistency.

\begin{acknowledgments}
One of the authors (A.V.Yu.) sincerely thanks the organizers and
participants of the International V.A.Fock School for Advances in
Physics 2002 (IFSAP-2002), where this paper was presented.
\end{acknowledgments}


\centerline{\bf REFERENCES} \noindent \begin{enumerate}


\bibliography{apssamp}

\bibitem{1} A. H. Guth, "The Inflationary Universe: A Possible
Solution To The Horizon And Flatness Problems," Phys. Rev. D 23,
347 (1981).

\bibitem{2} A. D. Linde, "A New Inflationary Universe Scenario: A
Possible Solution Of The Horizon, Flatness, Homogeneity, Isotropy
And Primordial Monopole Problems," Phys. Lett. B 108, 389 (1982);
A. Albrecht and P. J. Steinhardt, "Cosmology For Grand Unifed
Theories With Radiatively Induced Symmetry Breaking," Phys. Rev.
Lett. 48, 1220 (1982).

\bibitem{3} A.D. Linde, "Chaotic Inflation," Phys. Lett. 129B, 177
(1983).

\bibitem{4} S. Perlmutter et al. [Supernova Cosmology Project
Collaboration], Astrophys. J. 517, 565 (1999)
[arXiv:astro-ph/9812133]; A. G. Riess et al. [Supernova Search
Team Collaboration], Cosmological Constant," Astron. J. 116, 1009
(1998) [arXiv:astro-ph/9805201]; P. M. Garnavich et al.,
Astrophys. J. 509, 74 (1998) [arXiv:astro-ph/9806396].

\bibitem{5} T.P. J. Steinhardt, L. M. Wang and I. Zlatev, Phys. Rev. D
59, 123504 (1999) [arXiv:astro-ph/9812313]; B. Ratra and P. J.
Peebles, "Cosmological Consequences Of
 A Rolling Homogeneous Scalar Field,"
Phys. Rev. D 37, 3406 (1988); I. Zlatev, L. M. Wang and P. J.
Steinhardt, "Quintessence, Cosmic Coincidence, and the
Cosmological Constant," Phys. Rev. Lett. 82, 896 (1999)
[arXiv:astro-ph/9807002].


\bibitem{6}  Justin Khoury, Burt A. Ovrut, Paul J. Steinhardt and Neil
Turok. "The ekpyrotic universe: colliding branes and the origin of
the hot big bang." Phys. Rev. D64: 123522, 2001.
[arXiv:hep-th/0103239].

\bibitem{7} P. J. Steinhardt and N. Turok, "Cosmic evolution in a cyclic
universe," Phys. Rev. D 65, 126003 (2002) [arXiv:hep-th/0111098];
P. J. Steinhardt and N. Turok, "A cyclic model of the universe,"
[arXiv:hep-th/0111030]; P. J. Steinhardt and N. Turok, "Is Vacuum
Decay Significant in Ekpyrotic and Cyclic Models?"
[arXiv:astro-ph/0112537].

\bibitem{8} R. Dick, "Brane worlds", Class. and Quantum Grav., V. 18, N.
7, R1-24 (2001) [arXiv:hep-th/0105320].

\bibitem{9} R. Kallosh, L. Kofman and A. Linde, "Pyrotechnic universe".
Phys. Rev. D64: 123523, 2001. [arXiv:hep-th/0104073]; D. H. Lyth,
"The failure of cosmological perturbation theory in the new
ekpyrotic scenario," [arXiv:hep-ph/0110007];R. Brandenberger and
F. Finelli, "On the spectrum of fluctuations in an effective field
theory of the ekpyrotic universe," arXiv:hep-th/0109004.

\bibitem{10} Yu. V. Baryshev, "Conceptual Problems Of Fractal Cosmology",
Astronomical and Astrophysical Transaction, Vol. 19, pp. 417-435
(2000).

\bibitem{11} G. Felder, A. Frolov, L. Kofman and  A. Linde, "Cosmology
With Negative Potentials" [arXiv: hep- th/ 0202017]


\bibitem{12} R. Kallosh, A. D. Linde, S. Prokushkin and M. Shmakova,
"Gauged supergravities, de Sitter space and cosmology," Phys. Rev.
D 65, 105016 (2002) [arXiv:hep-th/0110089]; A. Linde, "Fast-roll
inflation", JHEP 0111, 052 (2001) [arXiv:hep-th/0110195].

\bibitem{13} J. D. Barrow, "Exact Inflationary Universes With Potential
Minima", Phys. Rev. D 49 3055 (1994); R. Maartens, D. R. Taylor
and Roussos, "Exact Inflationary Cosmologies With Exit", Phys.
Rev. D 52 3358 (1995);); V. M. Zhuravlev, S. V. Chervon, and V. K.
Shchigolev, JETP, 87, 223 (1998); S. V. Chervon,V. M. Zhuravlev,
and V. K. Shchigolev, Phys. Lett. B, 398, 269 (1997); J. D.
Barrow, Phys. Lett. B, 235,40 (1990); G. F. R. Ellis and M. S.
Madsen, Class. Q. Grav., 8, 667 (1991); J. E. Lidsey, Class. Q.
Grav., 8, 923 (1991);J. D. Barrow and P. Saich, Class. Q. Grav.,
10, 279 (1993); P. Parson and J. D. Barrow, Class. Q. Grav., 12,
1715 (1995).

\bibitem{14} A. D. Linde, Phys. Lett. B, 200, 272 (1988); Particle
Physics and In?ationary Cosmology, Harwood, Chur, Switzerland
(1990).

\bibitem{15} V. M. Zhuravlev and S. V. Chervon, JETP, 91, 227 (2000).

\bibitem{16}  A.V. Yurov, "Exact Inflationary Cosmologies With Exit: From
An Inflaton Complex Field To An Anti-Inflaton One", Class. Quantum
Grav., 18, 3753 (2001).

\bibitem{17} A. Kamenshchik, U. Moschella and V. Pasquier, "An
Alternative To Quintessence" [arXiv: gr-qc/0103004].


\bibitem{18}  A. D. Linde,"Hybrid inflation," Phys. Rev. D 49, 748 (1994)
[arXiv:astro-ph/9307002].

\bibitem{19}    V. B. Matveev V B and Salle M A \rm\, 1991 {\em Darboux
Transformation and
    Solitons} (Berlin--Heidelberg: Springer Verlag)

\bibitem{20}  J. G. Darboux, {\em C.R.Acad.Sci.,Paris 94} p.1456 (1882).

\bibitem{21}  M. M. Crum, "Associated Sturm-Liouville equations", {\em
Quart. J. Math. Oxford 6 2} p. 121 (1955).

\bibitem{22} F. J. Tipler and J. Graber, "Closed Universes With Black
Holes But No Event Horizons As a Solution to the Black Hole
Information Problem", [arXiv:gr- qc/ 0003082].

\bibitem{23} V. E. Adler, Theor. Math. Phys., 101, No. 3, 1381 (1995).

\bibitem{24} A. P. Veselov and A. B. Shabat, Funct. Anal. Appl., 27, No.
2, 81 (1993).

\bibitem{25} A. D. Linde, ``The Universe Multiplication And The
Cosmological Constant Problem,'' Phys. Lett. B 200, 272  (1988);
A.D. Linde, Particle Physics and Inflationary Cosmology (Harwood,
Chur, Switzer-land, 1990).

\bibitem{26} M. Tegmark,  "Is 'the theory of everything' merely the
ultimate ensemble theory?", Annals Phys. 270, 1 (1998)
[arXiv:gr-qc/9704009].

\bibitem{27} H. B. Nielsen and C. Froggatt, "Influence from the Future"
[arXiv: hep- ph/ 9607375]


\end{enumerate}
\end{document}